\documentclass{book}

\usepackage{piers}
\usepackage{amsmath}
\usepackage{amsthm}
\usepackage{amssymb}
\usepackage{mathtools}
\usepackage[utf8]{inputenc}
\usepackage{graphicx}
\usepackage{graphpap}

\newenvironment{bulletlist}{\begin{list}{$\bullet$}
{\setlength{\itemsep}{0em}
\setlength{\labelwidth}{15em} 
\setlength{\itemindent}{0 em}
\setlength{\parskip}{0em}
\setlength{\labelsep}{0.5em}
\setlength{\itemsep}{0pt}
     \setlength{\parsep}{1pt}
     \setlength{\topsep}{3pt}
     \setlength{\partopsep}{0pt}
     \setlength{\leftmargin}{1.5em}
     \setlength{\labelwidth}{1em}
     \setlength{\labelsep}{0.5em} }}
{\end{list}}

\def\multi#1{{\begin{tabular}[t]{@{}l@{}}#1\end{tabular}}}

\def\norm#1{\left|#1\right|}
\def\innerprod(#1,#2){{\left<#1\,,\,#2\right>}}
\def\Set#1{{\left\{#1\right\}}}
\def\qquadtext#1{\qquad\textup{#1}\qquad}
\def\qquadand{\qquadtext{and}}
\def\quadtext#1{\quad\textup{#1}\quad}
\def\quadand{\quadtext{and}}
\def\pfrac#1#2{\frac{\partial #1}{\partial #2}}
\def\ppfrac#1#2{\frac{\partial^2 #1}{\partial {#2}^2}}

\def\dfrac#1#2{\frac{d #1}{d #2}}

\def\Ve{{\boldsymbol e}}

\def\VE{{\boldsymbol E}}
\def\VB{{\boldsymbol B}}
\def\VD{{\boldsymbol D}}
\def\VH{{\boldsymbol H}}

\def\Vx{{\boldsymbol x}}

\def\VP{{\boldsymbol P}}

\def\FT#1{\widetilde #1}
\def\FTt#1{\hat{#1}}
\def\FtE{{\hat{E}}}
\def\FtP{{\hat{P}}}

\def\tE{{\widetilde{E}}}
\def\tP{{\widetilde{P}}}
\def\tL{{\widetilde{L}}}

\def\twopii{2\pi i}

\def\regL{{\mathbf L}}
\def\regR{{\mathbf R}}
\def\regmu{{\boldsymbol\mu}}

\def\Lag{{\cal L}}
\def\Act{{\cal S}}

\def\Podd{P^{(\textup{o})}}
\def\Pevn{P^{(\textup{e})}}
\def\omegaodd{\omega^{(\textup{o})}}
\def\omegaevn{\omega^{(\textup{e})}}
\def\EQ{{\cal Q}}

\author{Jonathan Gratus and Matthew McCormack}

\begin{document}
\title{Inhomogeneous Spatially Dispersive Electromagnetic Media}
\maketitle

\author      {Jonathan Gratus}
\affiliation {Lancaster University and the Cockcroft Institute}
\address     {Physics Department, Lancaster University}
\city        {Lancaster}
\postalcode  {LA1 4YB}
\country     {UK}
\phone       {+44 (0)1524 594980}    
\fax         {+44 (0)1524 844037}    
\email       {j.gratus@lancaster.ac.uk}  
\misc        { }  
\nomakeauthor

\author      {Matthew McCormack}
\affiliation {Lancaster University and the Cockcroft Institute}
\address     {Physics Department, Lancaster University}
\city        {Lancaster}
\postalcode  {LA1 4YB}
\country     {UK}
\phone       {}    
\fax         {}    
\email       {m.maccormack@lancaster.ac.uk}  
\misc        { }  
\nomakeauthor

\begin{authors}

{\bf Jonathan Gratus}$^{1}$, {\bf and Matthew McCormack}$^{1}$\\
\medskip
$^{1}$Physics Department, Lancaster University, LA1 4YB, UK and\\
The Cockcroft Institute, Sci-Tech Daresbury,
Keckwick Lane,
Daresbury,
Warrington,
WA4 4AD, UK
\end{authors}

\begin{paper}

\begin{piersabstract}
  Two key types of inhomogeneous spatially dispersive media are
  described, both based on a spatially dispersive generalisation of
  the single resonance model of permittivity. The boundary
  conditions for two such media with different properties are
  investigated using Lagrangian and distributional methods. Wave
  packet solutions to Maxwell's equations, where the permittivity
  varies and is periodic in the medium, are then found. 
\end{piersabstract}

\psection{Introduction} 

All media are, at least to some extent, both temporally and
spatially dispersive
\cite{agranovich1967spatial,agarwal1974electromagnetic,%
  belov2003strong,gratus2011covariant,gratus2010covariant}.  A
temporally dispersive medium takes time to respond to an
electromagnetic signal. A spatially dispersive medium responds not only to a
signal at a particular point, but to signals in the neighbourhood of
that point.  Likewise all media are inhomogeneous, both on the
macroscopic scale due to the finite nature of any sample of material,
and on the microscopic scale.

Compact, high gradient, accelerators have a wide
range of applications in academia, industry, energy and health.  A
dielectric wakefield accelerator\cite{rosenzweig:364} uses electrons
to create a field in a dielectric which in turn accelerates further
electrons.
Using a spatially dispersive dielectric with a periodic inhomogeneity,
requires a knowledge about which electromagnetic fields propagate in such
dielectric and how the fields pass through the vacuum-dielectric
boundary.

For this article we will assume that there is a linear constitutive
relationship between the polarization  field
$\VP(t,\Vx)=\VD(t,\Vx)-\VE(t,\Vx)$ and electric field
$\VE(t,\Vx)$. All
media respond linearly for sufficiently small electromagnetic fields
and ultimately all media, including the vacuum, are non linear for
sufficiently high fields. To simplify the analysis we make the
following assumptions. 
\begin{bulletlist}
\item
There is no magnetization so that $\VH=\VB$.
\item
All fields are functions of time $t$ and one spatial coordinate
$x=x_1$, thus independent of $x_2,x_3$.
In frequency domain,
$k_2=k_3=0$ and we set $k=k_1$.
\item The electric and polarization fields are transverse so that
  $E_1(t,x)=0$, $P_1(t,x)=0$ and $B_1(t,x)=0$.  This assumption
  automatically satisfies the two non-dynamic source free Maxwell's 
  equations.
\item
We choose a linearly polarized wave so that
in the $(\Ve_1,\Ve_2,\Ve_3)$ frame
$\VE(t,x)=E(t,x)\Ve_2$,
$\VP(t,x)=P(t,x)\Ve_2$ and
$\VB(t,x)=B(t,x)\Ve_3$
\end{bulletlist}

The relationship between $P(t,x)$ and $E(t,x)$ analysed here is a
generalisation of the single resonance model of permittivity.
The first generalisation is a simple extension to make the medium
spatially dispersive, achieved by introducing a finite propagation
speed $\beta$,
\begin{align}
\FT{P}(\omega,k)
=
\frac{\FT{E}(\omega,k)}
{\displaystyle
\gamma^2\omega^2 - 2i\gamma\lambda\omega +
  (\alpha^2-\lambda^2)-\beta^2 \norm{k}^2}
\label{Intro_homo_ft_CR}
\end{align}
where $\FT{P}(\omega,k) 
= \int_{-\infty}^{\infty} \int_{-\infty}^{\infty} e^{-\twopii(\omega t+ k x)}
P(t,x) dt\, dx
$ is the Fourier transform of $P(t,x)$.

The second generalisation is to allow the quantities
$\gamma,\lambda,\alpha$ and $\beta$ to depend on position $x$. 
To do this we replace the frequency-wave number relation
(\ref{Intro_homo_ft_CR}) with the PDE in space and time given by
\begin{align}
-
\frac{\gamma(x)^2}{(2\pi)^2} \ppfrac{P}{t}
+
\frac{2\lambda(x)\gamma(x)}{2\pi}\pfrac{P}{t}
+
\big(\alpha(x)^2-\lambda(x)^2\big) P
+
\frac{\beta(x)^2}{(2\pi)^2}\ppfrac{P}{x}
=
E
\label{Intr_P_PDE}
\end{align}
It is easy to see that if $\gamma,\lambda,\alpha$ and $\beta$ are
constants then the Fourier transform of (\ref{Intr_P_PDE})
reproduces (\ref{Intro_homo_ft_CR}).
We consider two types of
inhomogeneity.

In section \ref{ch_Bdd} we consider a simple boundary between two
homogeneous regions.  For spatially dispersive media the standard
boundary conditions for Maxwell's equations are insufficient to
completely specify the solutions for outgoing waves in terms of the
incoming waves. The additional equations are called ``additional
boundary conditions'' (ABC) and have often drawn
controversy\cite{pekar1958theory,zeyher1972spatial,henneberger1998additional}.
For the boundary between a spatially dispersive medium and the vacuum
the standard ABC are given by Pekar\cite{pekar1958theory}. In this
article we consider the boundary conditions between two spatially
dispersive regions. We consider two methods for deriving the boundary
conditions for Maxwell's equations for non-spatially dispersive media:
One is the distributional or pill box method, the other is to use a
Lagrangian method to derive {\em natural} boundary conditions. In
the case of non-spatially dispersive media, the two methods yield
identical boundary conditions. By contrast, if the propagation speed
$\beta$ in (\ref{Intr_P_PDE}) is different in the two regions,
then the pill box boundary conditions and the natural boundary
conditions will differ. In both cases the boundary conditions will
reduce to Pekar's boundary condition in the limit where $\beta\to0$ in
one of the regions.

\begin{wrapfigure}{r}{0.50\textwidth}
\vspace{1em}
\setlength{\unitlength}{0.048\textwidth}
\begin{picture}(10,7)(-0.5,-0.4)
\put(-0.45,-0.7){\includegraphics[height=7.9\unitlength]{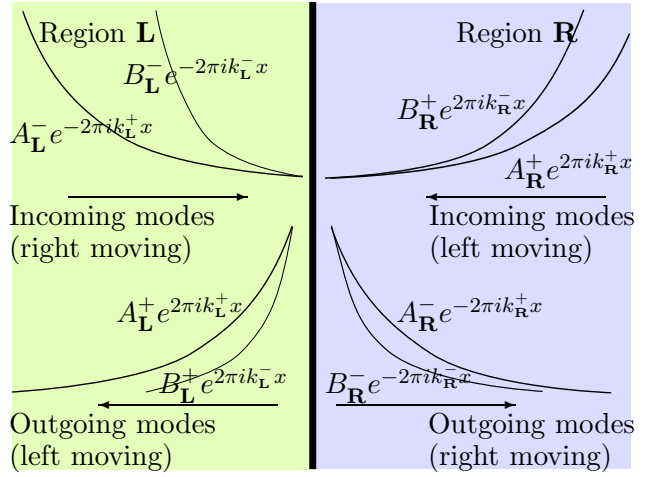}}
\put(0,6.5){Region $\regL$}
\put(7,6.5){Region $\regR$}
\put(-.5,4.8){$A^-_\regL e^{-\twopii k^{+}_\regL x}$} 
\put(1.4,5.8){$B^-_\regL e^{-\twopii k^{-}_\regL x}$} 
\put(.5,3.9){\vector(1,0){3}}
\put(-.5,3.8){\makebox(0,0)[tl]{\multi{Incoming  modes\\(right moving)}}}
\put(7.8,4.2){$A^+_\regR e^{\twopii k^{+}_\regR x}$} 
\put(6.,5.2){$B^+_\regR e^{\twopii k^{-}_\regR x}$} 
\put(9.5,3.9){\vector(-1,0){3}}
\put(10,3.8){\makebox(0,0)[tr]{\multi{Incoming modes\\(left moving)}}}

\put(1.3,1.8){$A^+_\regL e^{\twopii k^{+}_\regL x}$} 
\put(2.0,.6){$B^+_\regL e^{\twopii k^{-}_\regL x}$} 
\put(4,0.4){\vector(-1,0){3}}
\put(-.5,0.3){\makebox(0,0)[tl]{\multi{Outgoing modes\\(left moving)}}}

\put(6,1.8){$A^-_\regR e^{-\twopii k^{+}_\regR x}$} 
\put(4.8,0.6){$B^-_\regR e^{-\twopii k^{-}_\regR x}$} 
\put(5,0.4){\vector(1,0){3}}
\put(10,0.3){\makebox(0,0)[tr]{\multi{Outgoing modes\\(right moving)}}}
\end{picture}
\caption{Incoming and outgoing modes, given in (\ref{inhom_gen_soln_E}).}
\label{fig_In_Out}
\vspace{-1.5em}
\end{wrapfigure}

In section \ref{ch_Per} we consider a periodic structure where $\alpha$,
$\lambda$ and $\gamma$ are periodic with the same period. We assume
that these quantities take the form of a constant term plus a small
periodic inhomogeneity, e.g. $\alpha(x)=\alpha_0+\alpha_1\cos
x$ where the $\alpha_1$ is small.  
The goal in this
section is to find solutions to the source free Maxwell
equations, i.e. the dispersion relations. However since the medium is
inhomogeneous it is not possible to find single mode solutions of the form
$e^{\twopii (\omega t + k x)}$ and is therefore necessary to
look for packet solutions of the form $e^{\twopii\omega t}\FTt{P}(x)$.
In the following we present an analytic form for an approximate solution for
$\FTt{P}(x)$. In addition we present a numerical method for finding the
permitted frequencies and $\FTt{P}(x)$.

In section \ref{ch_Conc} we give a discussion of the implications
of this research.

\psection{Maxwell's equations and Stationary media} 
\label{ch_Max}

From $\VE(t,x)=E(t,x)\Ve_2$ etc., the source free dynamical Maxwell's
equations become
$E'=-\dot{B}$ and 
$B'=-(\dot{E}+\dot{P})$
which we can combine to give
${{E''}}=\ddot{E}+\ddot{P}$.
Taking the Fourier transform with respect
to $t$ 
gives
\vspace{-0.8em}
\begin{align}
(2\pi)^{-2}{\FtE''}=-\omega^2(\FtE+\FtP)
\label{Intr_Ej_FTt}
\end{align}
where $\FtP(\omega,x) = \int_{-\infty}^{\infty} e^{-\twopii \omega t}
P(t,x) dt$.  In most cases, in the following, we will not explicitly
write the $\omega$ argument in $\FtE$ and $\FtP$.

The Fourier transform of the constitutive relation (\ref{Intr_P_PDE}) is
\begin{align}
(2\pi)^{-2}{\beta^2(x)}{\FtP''}+L(x) \FtP = \FtE
\label{inhom_P_ODE}
\qquadtext{where}
L(x)
=
(\gamma(x)\omega+i\lambda(x))^2+\alpha(x)^2
\end{align}

\psection{Additional Boundary Conditions for spatially dispersive
  media}
\label{ch_Bdd}

In this section we set the media parameters
$\gamma(x),\alpha(x)$, $\beta(x)$, $\lambda(x)$ to be piecewise constant:
$\gamma(x)=\theta(-x)\gamma_\regL+\theta(x)\gamma_\regR$ etc., so that
$L(x)=\theta(-x)L_\regL+\theta(x)L_\regR$,
for media constants $\alpha_\regmu,\beta_\regmu,\lambda_\regmu\ge 0$
with $\regmu=\regL,\regR$. Here $\theta(x)$ is the Heaviside step function
$\theta(x)=0$ for $x<0$ and $\theta(x)=1$ for $x>0$.

Eliminating $\FtE$ from (\ref{Intr_Ej_FTt}) and (\ref{inhom_P_ODE})
we have a fourth order ODE for $\FtP(x)$. For each region this is
solved by
\begin{align}
\FtE(x)
&=
A^+_\regmu e^{\twopii k^+_\regmu x}
+
A^-_\regmu e^{-\twopii k^+_\regmu x}
+
B^+_\regmu e^{\twopii k^-_\regmu x}
+
B^-_\regmu e^{-\twopii k^-_\regmu x}
\label{inhom_gen_soln_E}
\end{align}
where
\vspace{-0.8em}
\begin{align}
k^{\pm}_\regmu=\frac{\sqrt{
\beta_\regmu^2\omega^2 + L_\regmu \pm
\sqrt{ (\beta_\regmu^2\omega^2-L_\regmu)^2 
+ 4\beta_\regmu^2\omega^2 } }}
{\sqrt{2}\beta_\regmu}
\label{inhom_k_Soln}
\end{align}
Since $\lambda_\regmu>0$ the waves are damped in the direction of
motion, (see figure \ref{fig_In_Out}).
Maxwell's equations give us two boundary conditions\vspace{-0.8em}
\vspace{-0.3em}
\begin{align}
[\FtE]=0 \qquadand [{\FtE}']=0
\label{Bdd_Maxwell_E}
\end{align}
where $[\FtE]$ is the discontinuity 
$[\FtE]=\lim_{x\to 0+} \FtE(x) - \lim_{x\to 0-} \FtE(x)$.
However, we need two additional boundary conditions for $[\FtP]$ and
$[{\FtP'}]$. In the usual scattering problem we prescribe the
incoming wave amplitudes
$\Set{A^-_\regL,B^-_\regL,A^+_\regR,B^+_\regR}$ and we calculate the
outgoing wave amplitudes
$\Set{A^+_\regL,B^+_\regL,A^-_\regR,B^-_\regR}$.

\psubsection{Lagrangian formulation of boundary conditions}
\label{ch_BddL}

Due to the damping term, it is non trivial to formulate a Lagrangian
which gives rise to both Maxwell's equations and the constitutive
relation (\ref{Intr_P_PDE}). However since we are interested in the
boundary conditions, is it sufficient to use the Fourier transform
equations (\ref{Intr_Ej_FTt}) and (\ref{inhom_P_ODE}). These can be
derived by varying the action
\begin{align}
\Act[\FtE,\FtP]
=
\int \Lag(\FtE,\FtE',\FtP,\FtP',x) \,dx
\label{BddL_Action}
\end{align}
where\vspace{-1.4em}
\begin{align}
\Lag(\FtE,\FtE',\FtP,\FtP',x)
=
\frac12
\Big(
\frac{(\FtE')^2}{(2\pi)^2\omega^2}- \FtE^2 
+\frac{\beta(x)^2}{(2\pi)^2} (\FtP')^2 - {L(x)} \FtP^2
\Big)
- \FtE\FtP
\label{BddL_Lag}
\end{align}
Varying (\ref{BddL_Action}) with respect to $\FtE$ and $\FtP$ away
from the boundary yields (\ref{Intr_Ej_FTt}) and (\ref{inhom_P_ODE})
respectively.  In order to obtain the boundary conditions we must
consider variations with support which includes the boundary $x=0$. It
is necessary to assume $\FtE$ and $\FtP$ are continuous, i.e.
\begin{align}
[\FtE]=0
\qquadand
[\FtP]=0
\label{BddL_P_cts}
\end{align}
Varying (\ref{BddL_Action}) with respect to $\FtP$ then gives
\begin{align*}
\delta_\FtP \Act
&=
\int_{-\infty}^{\infty} \delta_\FtP\Lag \,dx
=
\int_{-\infty}^{\infty} 
\Big(
\frac{\beta(x)^2}{(2\pi)^2 }\FtP'\,\delta\FtP' 
- {L(x)}\FtP\,\delta\FtP 
- \FtE\delta\FtP
\Big)
\,dx
\\&=
\int_{-\infty}^{0} \dfrac{}{x}
\Big(\frac{\beta(x)^2}{(2\pi)^2}\FtP'\,\delta\FtP\Big)
+
\int_{0}^{\infty} \dfrac{}{x}
\Big(\frac{\beta(x)^2}{(2\pi)^2}\FtP'\,\delta\FtP\Big)
=
-\bigg[\frac{\beta(x)^2}{(2\pi)^2}\FtP'\bigg]
\,\delta \FtP
\end{align*}
Since this vanishes for all variations $\delta\FtP$, one has
\begin{align}
[{\beta(x)^2 \FtP'}]=0
\label{BddL_Pp}
\end{align}
Similarly varying (\ref{BddL_Action}) with respect to $\FtE$ gives
$[\FtE']=0$.

In the limiting case $\beta_\regL\to0$,  the left hand region is
only temporally dispersive. One must make a choice about the behaviour
of $\Set{A_\regL^+,A_\regL^-,B_\regL^+,B_\regL^-}$ in this limit. For
a certain choice the boundary conditions
(\ref{Bdd_Maxwell_E}), (\ref{BddL_P_cts}) and (\ref{BddL_Pp}) reduce
to Pekar's ABC
\begin{align}
[\FtE]=0\,,\quad 
[\FtE']=0 \quadand
[\FtP]=0
\label{BddL_Pekar}
\end{align}

\psubsection{Distributional method of boundary conditions}

Given smooth functions $f_\regL(x)$ and $f_\regR(x)$ and setting 
$f(x)=\theta(-x) f_\regL(x)+\theta(x) f_\regR(x)$, then one has
$f''(x) = \delta'(x) [f]
+2\delta(x) [f']
+
\theta(-x) f''_\regL(,x) + \theta(x) f''_\regR(x)$.
Substituting this into (\ref{Intr_Ej_FTt}) and (\ref{inhom_P_ODE})
implies 
\begin{align}
[\FtE]=0\,,\quad
[\FtP]=0\,,\quad
[\FtE']=0\quadand
[\FtP']=0
\label{BddD_Bdd}
\end{align}
In the limit $\beta_\regL\to 0$, for appropriate choices, these again
reduce to Pekar's ABC (\ref{BddL_Pekar}).

\psection{Periodic Media} 
\label{ch_Per}

In this section we investigate media where the constitutive quantity
$L(x)$ in (\ref{inhom_P_ODE}) is periodic $L(x+1)=L(x)$ and $\beta$
is constant. We assume that the amplitude of the inhomogeneity is
dominated by the first mode, that is
\begin{align}
L(\omega,x)=L_0(\omega) + 2\Lambda(\omega)\cos(2\pi x)
\label{Per_L}
\end{align}
Taking the Fourier transforms of (\ref{Intr_Ej_FTt}) and
(\ref{inhom_P_ODE}) with respect to $x$ we get 
$(\omega^2 - k^2)\tE(k) = -\omega^2 \tP(k)$ and 
$- k^2 \beta^2 \tP + (\tL * \tP)(k) = \tE(k)$. Combining these into a
single equation gives
\begin{align}
(\tL * \tP)(k) = \left(\beta^2 k^2 - \frac{\omega^2}{\omega^2 - k^2}
\right) \tP(k)
\label{Per_Convolution_eq}
\end{align}
We look for periodic solutions of the form 
\vspace{-0.8em}
\begin{align}
\FtP(x) = \sum_{m=-\infty}^\infty P_{m} e^{2 \pi i m x}
\label{Per_P_anzats}
\end{align}
whose Fourier transform $\tP(k)$ consists of a series of delta
functions $\tP(k) = \sum_{m=-\infty}^\infty  P_{m} \delta(k-m)$.
Substituting (\ref{Per_P_anzats}) and (\ref{Per_L}) into
(\ref{Per_Convolution_eq}) yields the difference equation
\begin{align}
\Lambda P_{k-1} + f_k P_{k} + \Lambda P_{k+1} =
0
\label{Per_diff_eqn}
\qquadtext{where}
f_k 
= 
\frac{\beta^2 k^4 + \omega^2 + 
(\omega^2 -k^2)L_{0}(\omega) - \beta^2 \omega^2 k^2}{\omega^2 - k^2}
\end{align}
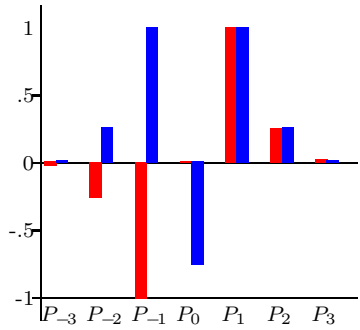
\begin{wrapfigure}{r}{0.3\textwidth}
\setlength{\unitlength}{0.6cm}
\begin{picture}(8.5,7.5)(-0.5,0)
\put(0,0){\footnotesize $P_{-\!3}$}
\put(1,0){\footnotesize $P_{-\!2}$}
\put(2,0){\footnotesize $P_{-\!1}$}
\put(3,0){\footnotesize $P_{0}$}
\put(4,0){\footnotesize $P_{1}$}
\put(5,0){\footnotesize $P_{2}$}
\put(6,0){\footnotesize $P_{3}$}

\put(-0.2,0.5){\line(1,0){7.2}}
\put(0,0){\line(0,1){7}}
\put(-0.2,3.5){\line(1,0){7.2}}
\put(0,2){\line(-1,0){0.2}}
\put(0,5){\line(-1,0){0.2}}
\put(-0.15,0.5){\makebox(0,0)[r]{\footnotesize -1}}
\put(-0.15,2.0){\makebox(0,0)[r]{\footnotesize -.5}}
\put(-0.15,3.5){\makebox(0,0)[r]{\footnotesize 0}}
\put(-0.15,5.0){\makebox(0,0)[r]{\footnotesize .5}}
\put(-0.15,6.5){\makebox(0,0)[r]{\footnotesize 1}}

\put(0,0.5){
\scalebox{1}[3]{\color{red}
\begin{picture}(7,3)(-3,-1)
\linethickness{0.25\unitlength}
\put(-3,0){\line(0,-1){0.0236108}}
\put(-2,0){\line(0,-1){0.252416}}
\put(-1,0){\line(0,-1){1}}
\put( 0,0){\line(0,1){0.01}}
\put( 1,0){\line(0,1){1}}
\put( 2,0){\line(0,1){0.252416}}
\put( 3,0){\line(0,1){0.0236108}}
\end{picture}
}}
\put(0.25,0.5){
\scalebox{1}[3]{\color{blue}
\begin{picture}(7,3)(-3,-1)
\linethickness{0.25\unitlength}
\put(-3,0){\line(0,1){0.02}}
\put(-2,0){\line(0,1){0.26}}
\put(-1,0){\line(0,1){1}}
\put( 0,0){\line(0,-1){0.75}}
\put( 1,0){\line(0,1){1}}
\put( 2,0){\line(0,1){0.26}}
\put( 3,0){\line(0,1){0.02}}
\end{picture}
}}
\end{picture}

\caption{$\Pevn$ (blue) and $\Podd$ (red) for $L_0\equiv 1$,
  $\Lambda\equiv 0.75$. In this case $\omegaevn=0.753i$ and $\omegaodd=0.399$.}
\label{fig_P}
\vspace{-0em}
\end{wrapfigure}

\noindent
Observe that having higher order modes in $L(\omega,x)$ will result in
more terms in (\ref{Per_diff_eqn}). We can trivially solve
(\ref{Per_diff_eqn}) simply by arbitrarily fixing $P_0$ and $P_1$ and
then using (\ref{Per_diff_eqn}) to solve for all subsequent
$P_k$. However, in general, this will lead to $P_k$'s which diverge
$|P_{k}|\to\infty$ as $k\to\pm\infty$. The Fourier transform of this
would therefore be a non-smooth wave which may not even be
continuous. Therefore for physical solutions, we demand that
$|P_{k}|\to 0$ as $k\to\pm\infty$. As we see below we can obtain approximate
analytic solutions for small $\Lambda$. In addition we also give a
numerical method for finding arbitrary solutions.

\psubsection{Approximate analytic wave packet solutions}

We have found two approximate analytic solutions in the case when
$\Lambda\ll L_0$, an even solution $(\omegaevn,\Pevn)$ and an odd
solution $(\omegaodd,\Podd)$.  Since these are approximate solutions, we
set the left hand side of (\ref{Per_diff_eqn}) to $\EQ_k$, that is
$\EQ_k=\Lambda P_{k-1} + f_k P_{k} + \Lambda P_{k+1} 
$
so that $\EQ_k=0$ is an exact solution to (\ref{Per_diff_eqn}). By contrast we
solve $\EQ_k=O(\Lambda^p)$ for some order $p$ which depends on $k$.

The even solution $\omega=\omegaevn$ is given by 
\begin{align}
\Big(f_1-\Lambda^2\Big(\frac{1}{f_2}+\frac{2}{f_0}\Big)\Big)
\Big|_{\omega=\omegaevn}=0
\label{Per_omegaevn}
\end{align}
$\Pevn$ is then given by
\begin{align}
\Pevn_{-1}=1\,,\quad
\Pevn_{0}=-\frac{2\Lambda}{f_0}\,,\quad
\Pevn_{1}=1
\quadand
\Pevn_{m} = \frac{(-\Lambda)^{|m|-1}}{\prod_{k=2}^{|m|} f_k} + 
O(\Lambda^{|m|+1})
\label{Per_Pevn}
\end{align}
By direct substitution into $\EQ_m$  shows 
$\EQ_0=0$, 
$\EQ_{\pm1}=O(\Lambda^4)$ and 
$\EQ_m=O(\Lambda^{|m|+1})$
for $|m|\ge2$.

The odd solution $\omega=\omegaodd$ is given by
\begin{align}
\Big(f_1-\frac{\Lambda^2}{f_2}\Big)
\Big|_{\omega=\omegaodd}=0
\label{Per_omegaodd}
\end{align}
and $\Podd_m$ by
\begin{align}
\Podd_{-1} = -1\,,\quad
\Podd_{0} = 0\,,\quad
\Podd_{1} = 1
\quadand
\Podd_{m} = \text{sign}(m) 
\frac{(-\Lambda)^{|m|-1}}{\prod_{k=2}^{|m|} f_k} + O(\Lambda^{|m|+1})
\label{Per_Podd}
\end{align}
Again $\EQ_0=0$, 
$\EQ_{\pm1}=O(\Lambda^4)$ and 
$\EQ_m=O(\Lambda^{|m|+1})$
for $|m|\ge2$.

Depending on how $L_0$ and $\Lambda$ depend on $\omega$ these are
often the lowest two modes. The shape of these modes is given in
figure \ref{fig_P}, using the numerical approach below. In this case
the even mode $(\omegaevn,\Pevn)$ cannot be supported by the medium
and is damped.

\psubsection{Numerical Approaches}

A numerical
approximation scheme, which is valid if $\Lambda\not\ll L_0$ and gives
packets in addition to
(\ref{Per_omegaevn}-\ref{Per_Pevn}),(\ref{Per_omegaodd}-\ref{Per_Podd}),
is as follows: Choose an integer $N\ge 2$. Then assume that
$P_m\approx 0$ for $|m|>N$ thus truncating the infinite set of
equation given by (\ref{Per_diff_eqn}) to a set of $2N+1$ linear
equations for $\Set{P_{-N},\ldots,P_{N}}$. Write this in matrix
language $M \underline b = \underline 0$ where $M$ is a
$(2N+1)\times(2N+1)$ matrix with $M_{k,k}=f_{k-N-1}$,
$M_{k,k-1}=M_{k,k+1}=\Lambda$ and $\underline b_k=P_{k-N-1}$.  Solve
$\det(M)=0$ to obtain values for $\omega$. The corresponding
null spaces give $P_m$.

\psection{Conclusion and Discussion}
\label{ch_Conc}

In this article we address two key
problems: What wave packets can propagate though a spatially
dispersive medium and how wave 
packets behave as they pass through a boundary between two media. 

We have given two methods of deriving the boundary conditions for a
junction between two spatially dispersive regions. These two methods
agree for the discontinuity of $\FtE$, $\FtE'$ and $\FtP$.  However
these two methods differ in the discontinuity of
$\FtP'$. Unfortunately in the limit $\beta_\regL\to 0$ both these
methods can be made to reduce to Pekar's ABC and so this cannot be
used to select one method above another. Thus one must decide, as part
of the model, which boundary condition to choose. This will depend on
the origin of the spatially dispersive single resonance constitutive
relation (\ref{Intro_homo_ft_CR}). If this comes from an underlying
model one may use that model to decide. On the other hand if it comes
from an action (\ref{BddL_Action}) or a PDE (\ref{Intr_P_PDE}) then
the corresponding boundary conditions can be applied. Alternatively,
this choice could be tested experimentally.

We have given approximate solutions to Maxwell's equations for a
periodic spatially dispersive medium. Since we have only given
two modes, it is natural explore the behaviour of a general mode.
From the numerical approach it appears that the higher frequency
resemble the case for homogeneous media. This is
currently being explored. 

\ack 

The authors are grateful for the support provided by STFC (the
Cockcroft Institute ST/G008248/1), EPSRC (the Alpha-X project
EP/J018171/1) and the Lancaster University Faculty of Science and
Technology studentship program. In addition the authors would like to
thank David Burton (Physics Department, Lancaster University) and
Adam Noble (Physics Department, Strathclyde University) for help in
preparing this article. 

\end{paper}


\end{document}